\documentclass[letterpaper,12pt]{article}   
\usepackage{osajnl2} 

\begin{document}

\title{ Study of wide-spectrum and high-resolution diffraction optical elements by stacks of binary phase gratings }

\author{I-Lin Ho$^{1,*}$, Wang-Yang Li$^{1}$}
\address{$^{1}$ChiMei Visual Technology Corporation, Tainan 741, Taiwan, R.O.C.}
\address{$^*$Corresponding author: sunta.ho@msa.hinet.net} 


\begin{abstract}
This work theoretically investigates wide-spectrum and high-resolution diffraction optical elements (DOE) that are made of stacks of low-resolution binary phase gratings, whereby the two-dimensional grids in different grating layers are arranged with specified displacements. We remodel the common Kinoform algorithm for this multi-scale architecture. Numerical computations show that, by increasing the number of stacking layers, the resolution of diffraction fields can be improved and that the stability of optical elements within broad spectrums is significantly enhanced.
Practical concern on largely increasing the number of grating layers are efficiency of the optical designs in theory and the manufacture of stacks of ultra-thin grating films.
\end{abstract}


\maketitle
\section{Introduction}
Modern day technology automation and advanced processing equipments for polymeric and composite materials have facilitated the production of more disciplinary and highly-creative products. The emergence of advanced coextrusion processing techniques especially allows for the fabrication of films composed of hundreds or thousands of multilayers with individual layer thicknesses down to a few tens of nanometers \cite{book1}. Using the unique characteristics of multilayered films has recently brought for diverse capabilities and new breakthroughs in material properties and multi-functional unitization. Examples include breathable films \cite{breath1}, optical gradient structures \cite{gradop1}, shape memory polymers \cite{shapeM1}, micro- and nano-fibers \cite{fiber1}, multilayer composites with brick-wall type microstructures \cite{brick1}, dual brightness enhancement film (DBEF) \cite{dbef1}, etc. This present work mainly explores multilayer architectures for optical holography.

Through computer generated holography (CGH) \cite{cgh1}, the field of optical holography has been used in many commercial applications, e.g. optical interconnection and diffraction optical elements (DOE) \cite{doe1}. Kinoform \cite{kinoform1} is one kind of phase hologram that manages wave fronts of an incident light by phase encoding, and can be operated with high diffraction efficiency. Many theoretical works also have reported on Kinoform optics under various approaches, e.g. simulated annealing method \cite{sa1,sa2,sa3} and direct binary search method \cite{dbs1}.

In this study, we apply multilayer architectures to study wide-spectrum and high-resolution Kinoforms, in case improving the physical characteristics of optical elements is not an easy-implemented option.
Similar to superresolution imaging \cite{imgreg1}, the high-resolution diffraction optical element (Kinoform) is realized by displacement stacks of low-resolution binary phase gratings (see Fig. \ref{fig1}). Moreover, the displacement stack correlates the high-resolution wave-front profile of devices with low-resolution binary grids, and makes the definitions of phase $\phi+2n\pi$ (n=0,1,2...) distinguishable and deterministic during optimization processes. All these features are proved to significantly enhance the stability of performance of current Kinoform architectures on large variations of incident wavelengths.

To carry out numerical calculations, we remodel the common Kinoform algorithm \cite{sa1} for the studied multi-scale architecture. Relevant codes by Matlab and Fortran-MEX can be downloaded online\cite{code1}. It is emphasized that, for optical elements with sub-wavelength resolutions, dominant interference effects should be considered in the optimization algorithms. Popular methods include the scalar diffraction model and effective medium model \cite{sdt1}. The present work only follows regular conditions for Kinoform. Another vector diffraction algorithm for phase holograms is preparing, which shall be reported for sub-wavelength optical devices in a future work.

\section{Geometric arrangements and algorithms}
\begin{figure}[tbp]
\begin{center}
\includegraphics[scale=0.8]{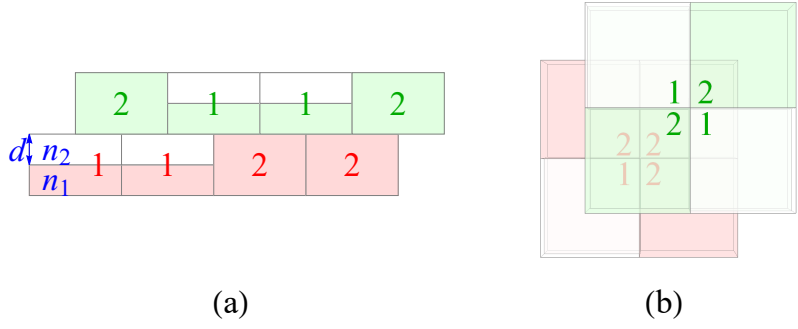}
\end{center}
\caption{Schematic diagram of displacement stacks of binary layers (a) in the side perspective and (b) in the top perspective, where displacement distance is set to be half of grid dimensions.} \label{fig1}
\end{figure}

\begin{figure}[tbp]
\begin{center}
\includegraphics[scale=0.7]{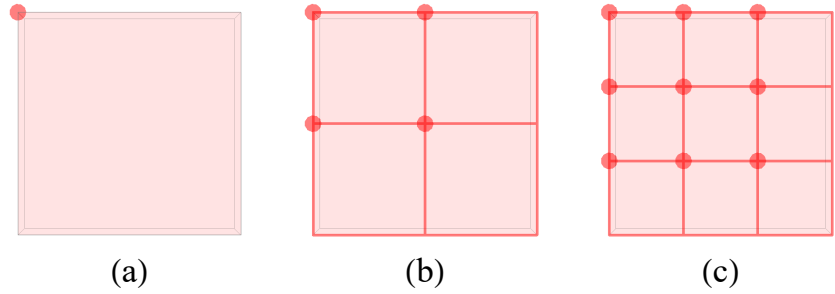}
\end{center}
\caption{Arrangements of displacement stacks of binary phase gratings for (a) original, (b) double, and (c) triple resolutions, which correspond to $1^{2}$, $2^{2}$, and $3^{2}$ displacement positions (red circles) within single grid (light-red square) of binary gratings, respectively.} \label{fig2}
\end{figure}

The optical elements presented herein are displacement stacks of optical layers with planar shifts at sub-grid accuracy. The optical layer is the binary phase grating that consists of two kinds of grids with optical path lengths: (i) $\phi\equiv n_{1}(t-d)+n_{2}d$ and (ii) $\phi+\lambda\Delta\Phi/2\pi$ (see Fig. \ref{fig1}). Here, $t$ is the thickness of the optical layer, $\lambda$ is the wavelength of the incident light, and $\Delta\Phi=2\pi(n_{1}-n_{2})d/\lambda$ defines the modulation phase of the binary grating. Figure \ref{fig1}(a) shows the stack of 2 binary grating layers with half-grid displacements from the side perspective, while Fig. \ref{fig1}(b) displays a double-resolution grid pattern from the top perspective. In this work, we adopt the arrangements of displacement stacks as Fig. \ref{fig2}, in order to extend to cases with arbitrarily high resolutions. Figure \ref{fig2} shows the arrangements for original, double, and triple resolutions of diffraction images, corresponding to $1^{2}$, $2^{2}$, and $3^{2}$ shift positions, respectively. Despite the enhancement of the resolution, we note that the reconstruction of images operates on optimizing low-resolution grids in all binary gratings. This implies that the high-resolution wave-front profile of devices is not operated independently as in conventional models, but correlates with the low-resolution grids of binary gratings. To treat the optimization of this multi-scale architecture, we remodel the common Kinoform algorithm \cite{sa1} here.

We denote the intensity profiles of the diffraction images by $I(x,y)$ on coordinate (x,y), and the phase profiles of $k_{th}$-binary grating by $\theta_{k}(u_{k},v_{k})$ on coordinate $(u_{k},v_{k})$. The high-resolution profile of the wave front of the optical element now is given by
\begin{eqnarray}
g\left( u,v\right)  &=&e^{i\sum_{\{k|u\in u_{k},v\in v_{k}\}}\theta
_{k}\left( u_{k},v_{k}\right) } \label{eq1} \\
\theta _{k}\left( u_{k},v_{k}\right)  &=&0,\Delta\Phi \label{eq2}
\end{eqnarray}
where the common factor $\phi$ has been omitted in Eq. (\ref{eq2}), since the global phase shift will not change physical observations.
The reconstructed diffraction images are $I(x,y)=|E(x,y)|^{2}=|F[g(u,v)]^{-1}|^{2}$. Here, $F[g(u,v)]^{-1}$ is the inverse Fourier transform of $g(u,v)$.
Following the iterative Fourier transform \cite{sa1} processes, one can obtain optimized phase profiles $\theta_{\{k\}}$ for all binary layers. To score the optimization algorithm, an error function \cite{sa1} is defined as the mean-square error between target image $I_{0}$ and reconstructed image $I$:
\begin{eqnarray}
f_{e}\left( \theta _{\{k\}}\right)  &=&\int \int \left\vert I_{0}(x,y)-\alpha I(x,y)\right\vert ^{2}dxdy \label{eq3} \\
\alpha  &=&\frac{\int \int I_{0}(x,y)dxdy}{\int \int I(x,y)dxdy} \label{eq4}
\end{eqnarray}

\section{Numerical simulations}

\begin{figure}[tbp]
\begin{center}
\includegraphics[scale=0.45]{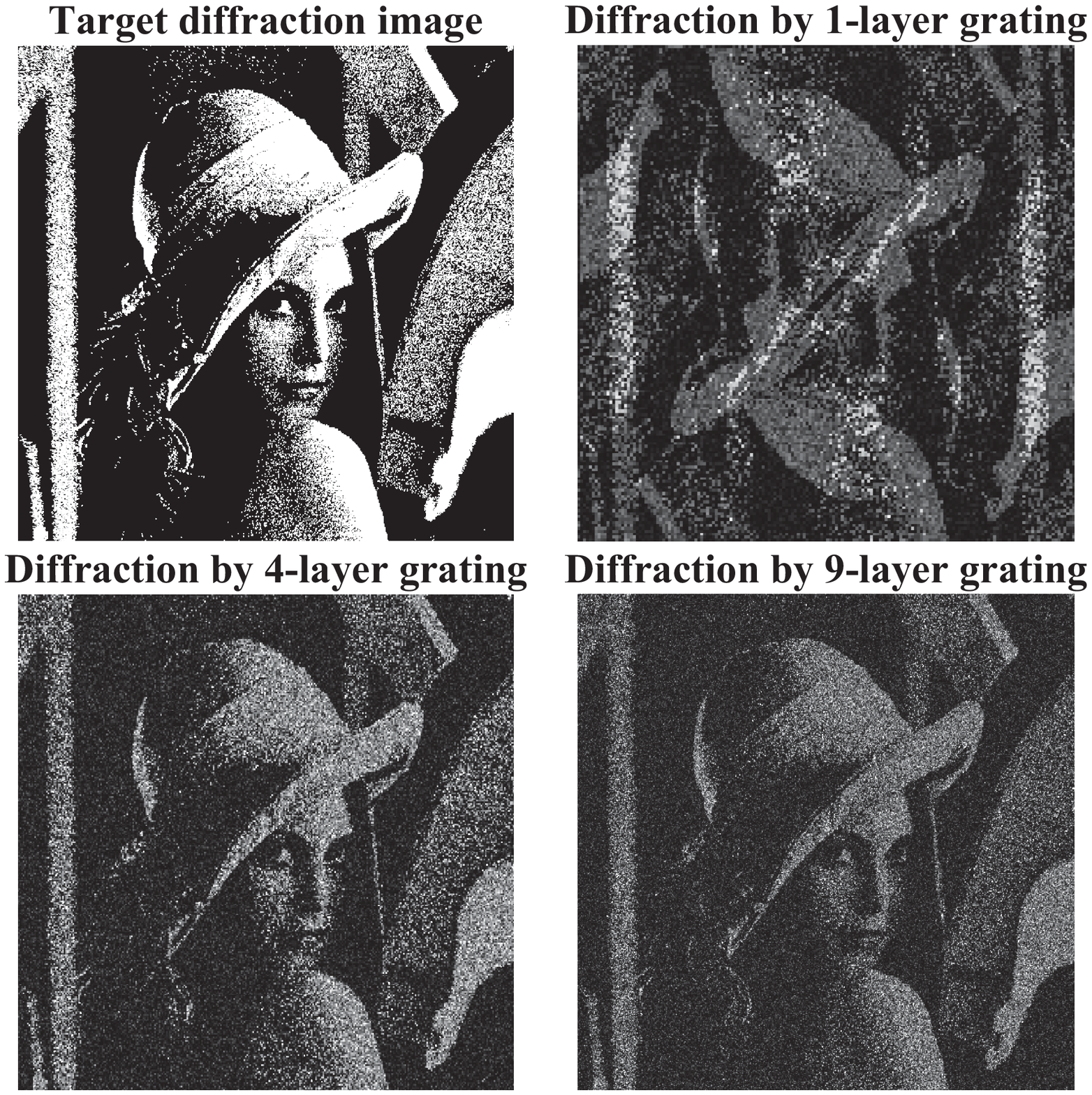}
\end{center}
\caption{Diffraction images of optical elements designed for different resolutions. (Upper left) Target of diffraction image, (upper right) diffraction image by single binary phase grating, (bottom left) double-resolution diffraction image by stacks of four binary phase gratings, and (bottom right) triple-resolution diffraction image by stacks of nine binary phase gratings.} \label{fig3}
\end{figure}

We first carry out digital simulations to investigate the performance of the present optical elements. The target diffraction image with 510$\times$510-pixels is depicted in the upper-left plot of Fig. \ref{fig3}. Binary phase gratings used in this work are set to have 170$\times$170-pixels (grids). The modulation phase $\Delta\Phi$ of binary gratings is set to be $\pi$ in the single-binary-layer case and $\pi/2$ in multilayer cases. In the case for using  single binary layer, the target image's resolution is reduced to that of the binary layer by simulating responses of sensors using the majority decision method. Numerical computations for the single binary layer shows a result similar to that by the conventional method, featuring image symmetry with respect to the origin (see the upper-right plot of Fig. \ref{fig3}). This origin symmetry is intrinsic for single binary phase grating. Equation (\ref{eqsym}) below gives an analytical inference.
\begin{eqnarray}
I\left( x,y\right)  &=&\left\vert E\left( x,y\right) \right\vert ^{2}\propto
\left\vert \left[\int\int g\left( u,v\right) e^{2\pi i(u\cdot x+v\cdot
y)}dudv\right] \right\vert ^{2}  \nonumber \\
&=&\left\vert \left[\int\int g^{\ast }\left( u,v\right) e^{-2\pi i\cdot
u\cdot x-2\pi i\cdot v\cdot y}dudv\right] ^{\ast }\right\vert ^{2}  \nonumber
\\
&\propto &\left\vert \left[ E\left( -x,-y\right) \right] ^{\ast }\right\vert
^{2}=I\left( -x,-y\right)   \label{eqsym}
\end{eqnarray}
Here, the condition for single binary phase grating, $g(u,v)=\{e^{0},e^{i\pi}\}=g^{*}(u,v)$, is applied for the derivation of the last-line formula. From Eq. (\ref{eqsym}), one knows that the diffraction intensity $I(x,y)$ is always the same with $I(-x,-y)$ for the single-binary-layer case, obeying origin symmetry. For the double-resolution design, we adopt the arrangement in Fig. \ref{fig2}(b) to stack four binary layers. The resolution of the target image is correspondingly treated by simulating sensor responses. The computed diffraction image is depicted in the bottom-left plot of Fig. \ref{fig3}, which inherits properties of four-level phase gratings and shows outlines closed to the target plot. For the triple-resolution condition, we adopt the arrangement in Fig. \ref{fig2}(c) to stack nine binary gratings. The computed diffraction image, in the bottom-right plot of Fig. \ref{fig3}, presents almost the same details with the target image. It is concluded that the multilayer architecture can improve the resolution of diffraction images by optimizing low-resolution grid profiles in all displaced binary gratings.

To investigate the performance of the optical element within broad spectrums, we study the deviations of diffraction images from the target one at different incident wavelengths. Here, the diffraction deviation is quantified by the error function in Eq. (\ref{eq3}). The variable of incidence wavelengths $\lambda$ is associated with the modulation phase $\Delta\Phi$ by the equation $\Delta\Phi\equiv2\pi\Delta n d/\lambda$, with $\Delta n=n_{1}-n_{2}$. We analyze four kinds of optical elements: (i) conventional single 4-level($0$,$0.5\pi$,$\pi$,$1.5\pi$) grating layer with 340$\times$340 pixels; (ii) stacking four binary gratings (170$\times$170 pixels/each), using the double-resolution scheme of Fig. \ref{fig2}(b); (iii) stacking eight binary gratings, two layers per position in the double-resolution scheme; and (iv) stacking twelve binary gratings, three layers per position in the double-resolution scheme.

The resolution of the target image is scaled as 340$\times$340 pixels to coincide with that of the four optical elements. It is noted that optical elements (i) and (ii) have maximal modulation phase $2\pi$ as in conventional designs \cite{sa1}. Optical elements (iii) and (iv), however, have maximal modulation phase $4\pi$ and $6\pi$, respectively. In contrast with conventional applications \cite{sa1}, the definitions of phase $\phi+2n\pi$ (n=0,1,2,...) in the studied multilayer architectures are distinguishable and deterministic during optimization processes, since the displacement stack correlates the high-resolution wave-front profile $g$ with the low-resolution binary grids $\theta_{\{k\}}$ as in Eq. (\ref{eq1}).

\begin{figure}[tbp]
\begin{center}
\includegraphics[scale=0.5]{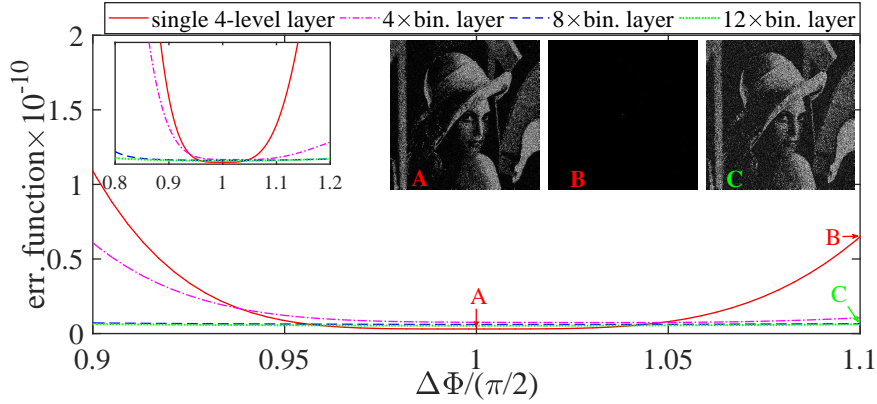}
\end{center}
\caption{Relations of diffraction deviations and incident wavelengths for (i) conventional 4-level phase grating, (ii) stacks of four binary phase gratings, (iii) stacks of eight binary phase gratings, and (iv) stacks of twelve binary phase gratings. The diffraction deviations are defined as an error function in the texts. The dependence of incidence wavelengths $\lambda$ is related to the modulation phase $\Delta\Phi\equiv\Delta n d/\lambda$ of the binary grating. } \label{fig4}
\end{figure}

Numerical results about the relation of diffraction deviations and incident wavelengths are shown in Fig. \ref{fig4}. Here, the conventional 4-level grating \cite{sa1} gives less diffraction deviation than others at the designed value $\Delta\Phi=\pi/2$ (see point A on the red-solid curve and the corresponding inset image A). As $\Delta\Phi$ is away from the designed value ($>|\pm5\%|$), diffractions of the conventional grating significantly deviate from the target profile and feature an image with overwhelming central zero-order intensity (see point B on the red-solid curve and the corresponding inset image B).  The 4$\times$binary-layer optical element presents variations of diffraction similar to that of the conventional case near the designed value, since it inherits properties of 4-level phase gratings.

We notice that there are three differences between the conventional grating and the 4$\times$binary-layer grating: (i) the performance of the conventional grating is better than that of the 4$\times$binary-layer grating on the designed condition $\Delta\Phi=\pi/2$; (ii) the decline in performances of the 4$\times$binary-layer grating is less than that from conventional grating as $\Delta\Phi$ is away from the designed condition; and (iii) deviations in performances of the 4$\times$binary-layer grating are obviously asymmetric in the case that $\Delta\Phi$ is away from the designed condition in increasing and decreasing directions.

\begin{figure}[tbp]
\begin{center}
\includegraphics[scale=0.6]{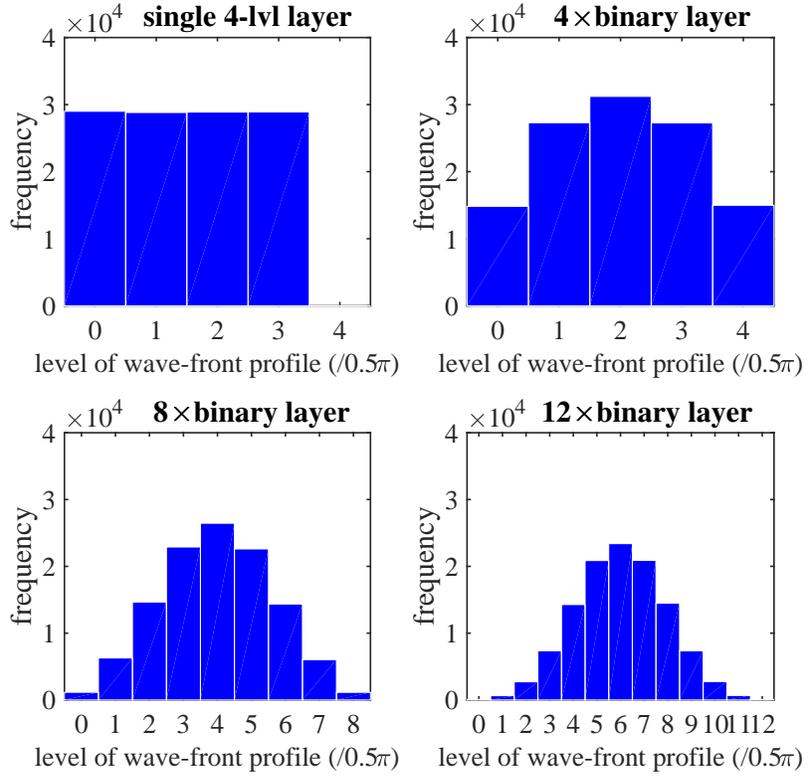}
\end{center}
\caption{Histogram of phase levels of wave-front profiles for optical elements using: (a) conventional 4-level phase grating, (b) displacement stacks of four binary gratings, (c) displacement stacks of eight binary gratings, and (d) displacement stacks of twelve binary gratings. } \label{fig5}
\end{figure}

\begin{figure}[tbp]
\begin{center}
\includegraphics[scale=0.60]{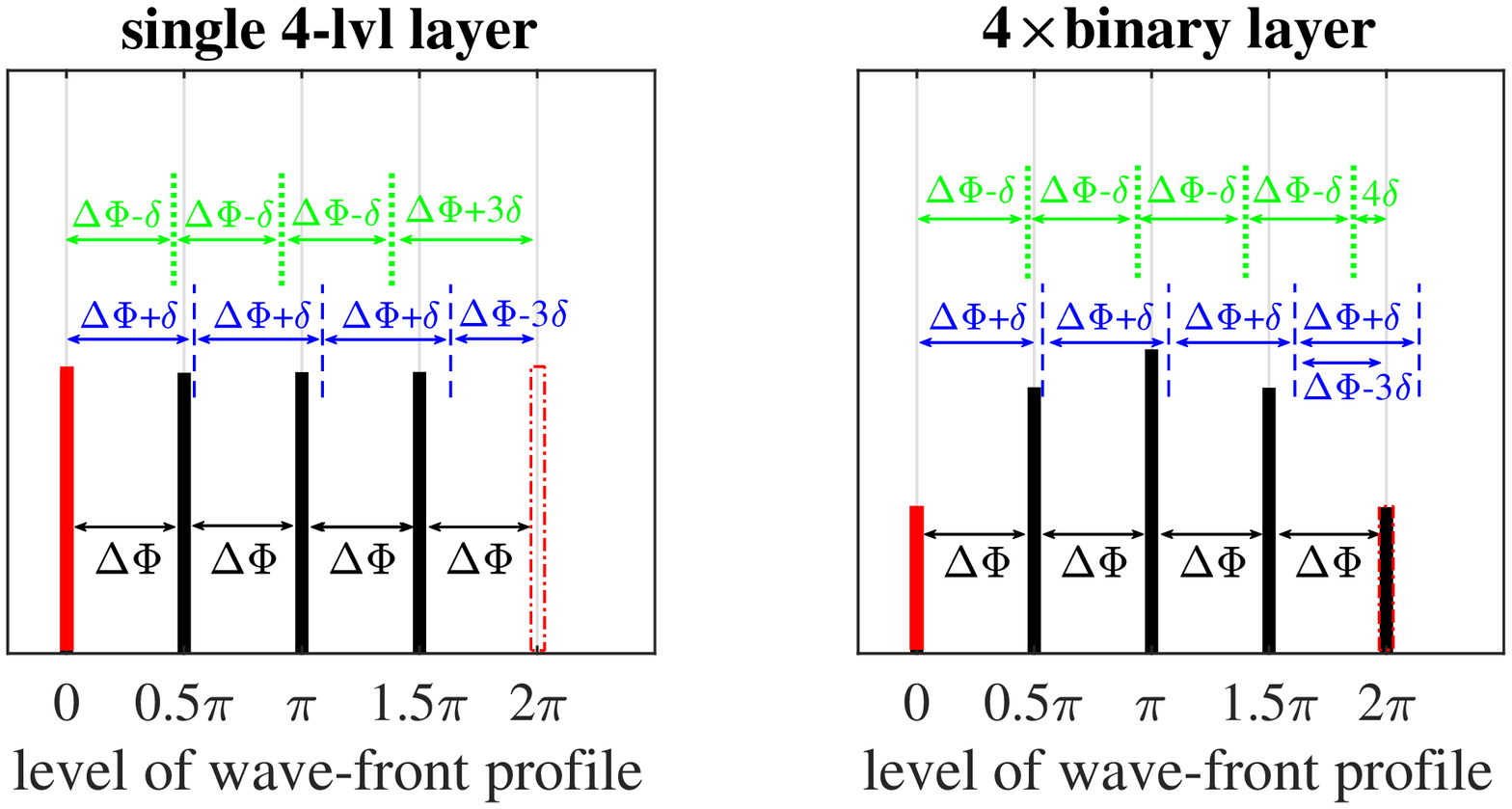}
\end{center}
\caption{Variations of histograms of wave-front profiles with respect to increasing and decreasing $\Delta\Phi\rightarrow\Delta\Phi\pm\delta$ in (a) the conventional 4-level grating and (b) the 4$\times$binary-layer optical element, as defined in Fig. \ref{fig5}(a-b). } \label{fig6}
\end{figure}

To understand these differences, we calculate histograms of phase levels of wave-front profiles for relevant optical elements in Fig. \ref{fig5}. The variations of levels of wave-front profiles with respect to increasing and decreasing $\Delta\Phi$ are shown in Fig. \ref{fig6}, in which the solid bars in Fig. \ref{fig6}(a-b) redraw the histograms of wave-front profiles for the conventional grating and the 4$\times$binary-layer grating as in Fig. \ref{fig5}(a-b). The red-dot-dash bars in Fig. \ref{fig6} represent the underlying/intrinsic modulation level owing to the $2\pi$-periodicity of phase.

We discuss some findings here:

(i) As indicated in Fig. \ref{fig5}(b-d), the optical elements by displacement stacks present Gaussian-type distributions on levels of wave-front profiles, in contrast with the uniform distribution for the conventional one in Fig. \ref{fig5}(a). This constraint about Gaussian-type distribution results from correlations between high-resolution wave-front pixels and low-resolution binary grids in the 4$\times$binary-layer grating, and reduces the degrees of freedom of the modulation phase during optimization processes. Consequently the performance of the conventional grating is better than that of the 4$\times$binary-layer grating on the optimized condition $\Delta\Phi=\pi/2$.

(ii) In Fig. \ref{fig6}(a), with increasing or decreasing $\Delta\Phi\rightarrow\Delta\Phi\pm\delta$, different phase intervals of wave-front levels do not equidistantly deviate for the conventional grating, especially the fourth interval ($\Delta\Phi\pm3\delta$). The abrupt (fourth) modulation interval significantly disturbs the regularity of constructive and destructive interference as well as the management of wave fronts of the incident light. For the 4$\times$binary-layer grating, since the fifth modulation level ($\Delta\Phi=2\pi$) exists, different modulation intervals can equidistantly deviate with varying $\Delta\Phi$ (see Fig. \ref{fig6}(b)). Hence, the performance of the 4$\times$binary-layer grating is better than that of the conventional grating at varying $\Delta\Phi$ away from the designed condition.

(iii) In Fig. \ref{fig6}(b), the underlying/intrinsic level (red-dot-dash bar) at $\Delta\Phi=2\pi$ also participates in the phase modulation. This level particularly contributes to asymmetric modulation intervals $\Delta\Phi-3\delta$ and $4\delta$ on increasing and decreasing $\Delta\Phi$, respectively. The asymmetry of these two modulation intervals implies distinct deviations of performances in the 4$\times$binary-layer grating when increasing and decreasing $\Delta\Phi$ (see the purple-dot-dash curve in Fig. \ref{fig4}).

For the optical elements with 8 binary layers and 12 binary layers in Fig. \ref{fig4}, the performances are much stable under largely varying $\Delta\Phi$ ($\pm10\%$, see the inset plot C in Fig. \ref{fig4}), or alternatively, within a broad incident spectrum.
It is inferred that optical elements with numerous displacement stacking layers decrease the populations of side levels (see Fig. \ref{fig5}(c-d)) and irregular modulation intervals at varying incident wavelengths. Moreover, spatial interferences among nearby pixels in the wave-front mask will seldom cross large phase intervals, owing to the non-trivial pixel correlations. All these factors benefit stable performances of the present Kinoform architectures on large variations of incident wavelengths.

\section{Conclusion}
This work studies multilayer architectures for diffraction optical elements, and presents relevant optimization algorithms.  Numerical computations show that high-resolution diffraction images can be realized by devices using displacement stacks of low-resolution binary phase gratings. In additions, with the number of binary layers $\ell$ up to $\ell\geq4\pi/\Delta\Phi$, this element presents a much more stable optical performance on variations ($<20\%$) of incident wavelengths.

\section{Acknowledgement}
This work was supported by ChiMei Visual Technology Corporation.

\end{document}